\documentclass[a4paper]{sig-alternate} 

\usepackage{times}
\usepackage{graphicx}
\usepackage{booktabs}
\usepackage{array}
\usepackage[utf8x]{inputenc}
\usepackage{cleveref}
\usepackage[numbers]{natbib}
\usepackage{amssymb}
\usepackage{tikz}
\usepackage{xspace}
\usepackage{url}
\usepackage{caption}\DeclareCaptionType{copyrightbox}\usepackage{subcaption} 

\newcommand{\alg}{reach-parameterized vector clocks\xspace}
\newcommand{\vclp}{VCLP\xspace} \newcommand{\prc}{precision-recall curve\xspace}
 
\newcommand{\election}{\texttt{election}\xspace}
\newcommand{\electionExt}{\texttt{election2}\xspace}
\newcommand{\olympics}{\texttt{olympics}\xspace}
\newcommand{\irvine}{\texttt{irvine}\xspace}
\newcommand{\studivz}{\texttt{studivz}\xspace}
\graphicspath{{figs/}}

\begin{document}
\title{Link Prediction with Social Vector Clocks} 

\numberofauthors{4}

\author{
\alignauthor
Conrad Lee\\
\affaddr{University College Dublin}\\
\mbox{\email{{\small conradlee@gmail.com}}}
\alignauthor
Bobo Nick\\
\affaddr{University of Konstanz}\\
\mbox{\email{{\small bobo.nick@uni-konstanz.de}}}
\and
\alignauthor
Ulrik Brandes\\
\affaddr{University of Konstanz}\\
\mbox{\email{{\small ulrik.brandes@uni-konstanz.de}}}
\alignauthor
P\'adraig Cunningham\\
\affaddr{University College Dublin}\\
\mbox{\email{{\small padraig.cunningham@ucd.ie}}}
}

\maketitle

\begin{abstract}
  State-of-the-art link prediction utilizes combinations of complex
  features derived from network panel data.  We here show that computationally
  less expensive features can achieve the same performance in the common
  scenario in which the data is available as a sequence of interactions.
  Our features are based on social vector clocks, an adaptation of the
  vector-clock concept introduced in distributed computing to social
  interaction networks.  In fact, our experiments suggest that by taking
  into account the order and spacing of interactions, social vector clocks
  exploit different aspects of link formation so that their combination
  with previous approaches yields the most accurate predictor to date.
\end{abstract}
\category{H.2.8}{Database Management: Database Applications --- Data Mining}

\section{Introduction}

Link prediction deals with predicting previously unobserved interactions
among actors in a network \cite{lc-lpfee-12}. Predictions are based on
the dynamic network of previously observed interactions, which is usually
made available in one of two forms: panel data or event data.  The former
refers to a sequence of complete network snapshots and typically
contains only coarse-grained temporal information.
Event data, on the other hand, consists of a finer-grained sequence of
single, time-stamped relational events, in which the exact minute or
second of each event is known.  Whereas panel data is often collected
by means of longitudinal surveys, event data is typically the outcome
of automated data collection, such as log files of e-mail, phone, or
Twitter communication.

It is possible to convert network event data into (interval-censored) network panel data;
indeed, \citet{ln-lppsn-03} did so in their seminal paper which
introduced the link prediction problem for social networks, and the
practice has become standard \citep{lc-lpfee-12,lz-lpcns-11}.  The
conversion is usually carried out by defining a sequence of time
slices and aggregating relational events within these time windows
into static (weighted) networks. At the expense of losing the ordering
and spacing of original events, this conversion allows one to employ
the large set of tools that have been developed for static and
longitudinal network analysis \citep{wf-snama-94,s-mlna-05}.

Whether the conversion from event to panel data is justifiable or not
depends crucially on mechanisms which drive tie formation in a given
network. For example, \citet{ln-lppsn-03} conducted experiments on future
interactions in large co-authorship networks. In this setting, the exact
sequence and spacing of publication dates can hardly be relevant because
publications dates are distorted anyway (backlogs, preprints, etc.);
aggregation of publication events on a coarser time-scale thus does not
appear to be problematic.

In other situations the fine-grained temporal information may be
highly relevant, making the conversion from event data to panel data
difficult to justify because it may destroy important patterns of
interaction; see \cite{hs-tn-12} for a recent review on general
temporal network approaches to exploit such information.  With regard
to the link prediction problem, if we are trying to foresee whether
node~A will send an email message to~B in the near future, for
example, then an extremely useful piece of information is whether~B
has \emph{recently} sent~A a message; if so, it is likely that~A will
respond to~B soon. This response-mechanism is known as
\emph{reciprocity}, and has been observed to be highly relevant for
predicting future events in social networks \citep{bls-saded-09}. By
aggregating communication events into cross-sectional graphs,
traditional link prediction schemes are generally prone to miss such
simple and useful mechanisms.

Here, we demonstrate that link predictors can indeed be made more
effective and efficient if they operate directly on appropriate
time-stamped dyadic communication data, and as a result can take
advantage of the information contained in the spacing and ordering of
relational events.  The approach we introduce is based on keeping
track of how out of date a node~A is with respect to another node~B with
respect to time-respecting information flow, and for doing so we
employ the concept of vector clocks.  Our results confirm that
dyadic features that exploit fine-grained temporal information
can be highly relevant for predicting which actors will
communicate for the first time in the near future, and are not limited to reciprocity.

The outline is as follows: first, we describe the types of data for
which we expect fine-grained temporal data to be relevant for link
prediction. Next, we review the link prediction problem for this type
of data, paying particular attention to supervised link prediction, a
framework that we employ here.  We then specify how a modified version
of \emph{social} vector clocks can be used as a supervised link
predictor, and proceed to evaluate this predictor.  We conclude with a
discussion of our results, as well as possible future work.


\section{Motivation} 
\label{sec:motivation}
\begin{figure}[t]
  \centering
  \includegraphics[width=.4\textwidth]{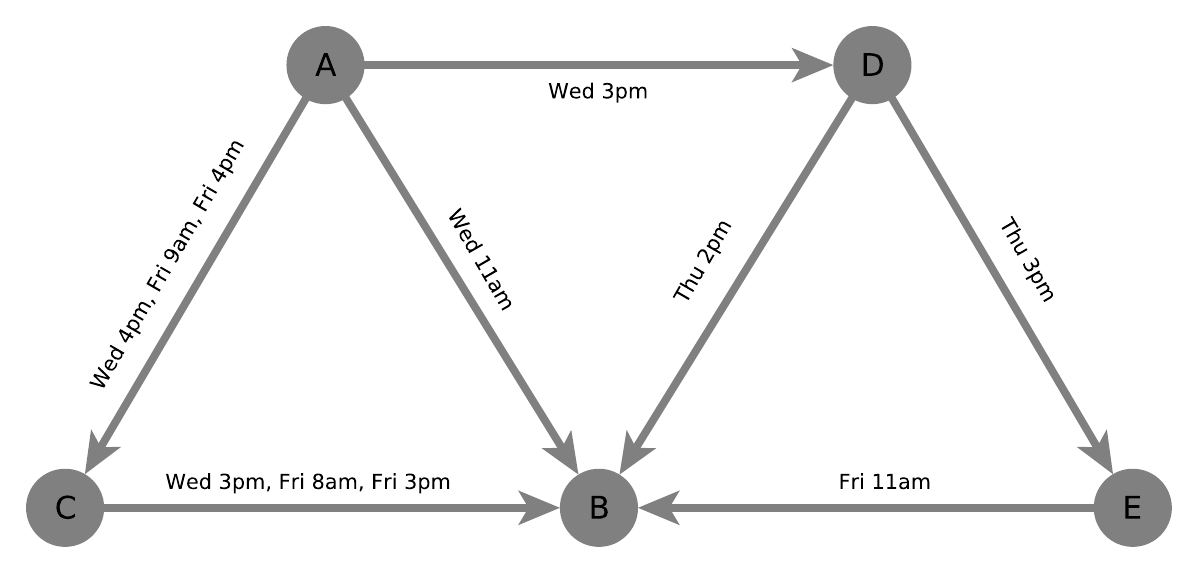}
  \caption{\label{fig:example} Illustrative example of social network data containing fine-grained information on dyadic communication events.
  Indirect information might flow on time-respecting paths, i.e.~along
  labels that respect the ordering of time. Adapted from \citep{kkw-sipscn-08}.}
\end{figure}

In the introduction, we stressed that the ordering and spacing of
communication events might contain valuable additional information
over and above the mere number of contacts between a pair of actors.
Consider the example in Figure~\ref{fig:example}, which depicts a
series of directed communication events, such as e-mail messages. Let
us imagine that the actor represented by node~A is in charge of
organizing a wellness weekend-trip for a group of friends, and that
she keeps changing her mind about when and where to go.  She finally
settles on a plan on Thursday at noon, and all of the subsequent
messages she sends out include the final trip details. We can ask:
which nodes can possibly know them, given the observed interactions?
Clearly, node~A communicated with node~C after she made up her mind on
Thursday, so node~C would have received the final information on
Friday at~9am. Because node~C subsequently sent node~B a message,
node~B could also have received the correct information.  On the other
hand, nodes~D and~E could not have received information from node~A
that is more recent than Wednesday at 3pm.

Two key and related concepts present in this example are
\emph{latency} and \emph{indirect updates}. We expect that groups of
people who coordinate some action, such as a wellness weekend-trip,
will need to synchronize their knowledge of certain key information
such as departure time and destination.  So in some
sense (which we will define formally in \Cref{sec:vector-clocks}) the
members of this group have a low latency with respect to each
other. Indirect updates, such as the one that made~B aware of~A's
latest trip plans, are an important mechanism for maintaining these
low latencies: even though~B did not have any direct message from~A
after she made up her mind, B~still got the latest plans indirectly
via~C.

This type of indirect communication is common in social systems:
consider the case of several adult siblings who communicate with each
other rarely, and more often communicate with their parents. In this
case, the siblings' information on each other can remain up to date
due to the central role of the parents, who provide the siblings with
an indirect means of communication. More generally, we observe that
gossip~-- the information exchanged when two people talk about an
absent third party~-- is a form of communication prevalent in society
and is in essence a form of indirect update.

The motivation behind our approach in this paper is that this
diffusion of information via indirect updates is common in many
social systems, and can be exploited to infer future direct
relations. In terms of the example above, we might predict that the
siblings are likely to communicate with each other because their
latencies with respect to each other remain low. However, the current
approach to link prediction throws away much of these temporal clues
by first converting the event stream into panel data.

In the datasets we analyze here, we also have reason to believe that
fine-grained diffusion patterns are relevant to link prediction.  For
instance, two of the datasets that we will use during our evaluation
contain sequences of micro-blogging events that come from Twitter.
\citet{bhmw-eiqit-11}, e.g., have found that word-of-mouth information
in Twitter spreads via many small cascades of tweets, mostly triggered
by ordinary individuals. These small chains of diffusion are exemplary
of the indirect updates we described above, and by considering the
details of how information spread, we may be able to infer which nodes
will come into direct contact in the future.  Detailed information on
the datasets we use in our final evaluation is given in
\Cref{datasets}.


\section{Link Prediction}
\label{sec:vclp}
In this section, we review the basics of link prediction.
In particular, we provide an overview of how machine learning models can be
used to combine multiple link predictors~-- a technique called supervised link
prediction. 

\subsection{The Problem and its Evaluation}
\label{sec:lp-evaluation}
Along the lines of its original formulation by
\citeauthor{ln-lppsn-03} \cite{ln-lppsn-03}, we formulate the link
prediction problem for dyadic event data as follows:
\begin{quote}
  Given a sequence of communication events in the form of (time,
  sender, receiver) tuples, predict which pairs of nodes who had no
  communication (i.e.~are disconnected) in the time interval
  $[t_0,t_1)$ will communicate (i.e.~become connected) in the time
  interval $[t_1,t_2)$.\footnote{In practice, the specification of a
    link prediction task involves more details, such as whether
    directed or undirected dyads are considered; we address these
    points in \Cref{eval-setup}.}
\end{quote}
An \emph{unsupervised link predictor} is a function which, given a
dyad (a pair of nodes) and the list of all previously occurring
events, returns a score, where a higher score indicates that an edge
is more likely to form in the dyad. \emph{Common neighbors} is an
example of an unsupervised link predictor: given a dyad~(A,B),
return the number of contacts shared by~A and~B. Although common
neighbors is simple, it is quite effective and many of the most
effective unsupervised link predictors (such as the similarity measure
originally proposed by Adamic and Adar \cite{aa-fnw-03}) are also
based on shared neighbors \cite{ln-lppsn-03,lz-lpcns-11}.

By running a link predictor on all dyads that are disconnected in the
interval $[t_0, t_1)$, one can rank all of the possible new links. To
evaluate a link predictor, we compare this ranking with the set of new
dyads that actually occur in the period $[t_1,t_2)$. In practice,
performance on the link-prediction task is often measured using ROC
curves or measures based on precision, but in their recent paper on
evaluation in the link prediction problem, \citet{lc-lpfee-12}
convincingly argue that due to the extreme class imbalance present in
the link prediction task, precision-recall curves are a more relevant
and less deceptive way to measure performance.  For that reason, here
we exclusively use precision-recall curves for our evaluation.

\subsection{Supervised link prediction}
\label{sec:supervised-lp}
As link prediction is fundamentally a binary classification problem,
it is natural to use the powerful binary classification models that
have been developed in machine learning. The primary advantage of this
approach is the ability to combine multiple unsupervised link
predictors into one joint prediction model. We will now provide a
brief overview of how supervised link prediction works. For an
in-depth discussion of supervised link prediction, see
\cite{ll-npmlp-10}.

\begin{figure}[h]
  \centering
  \includegraphics[width=0.98\linewidth]{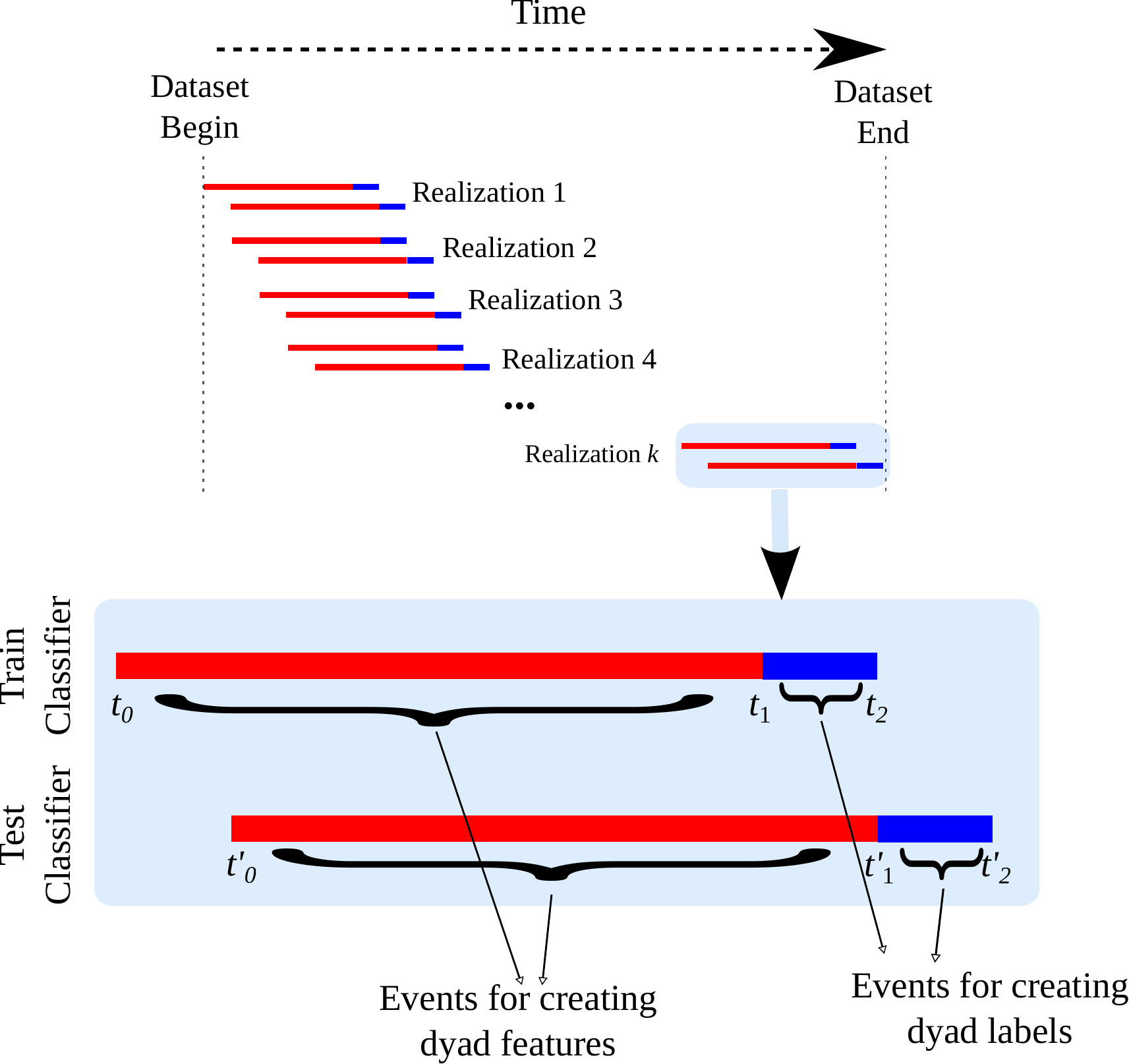}
  \caption{Framework for performing and evaluating supervised link
    prediction. Each dataset is split into several \emph{realizations}; each
    realization, in turn, is split into intervals to train and test a
    classifier.}
  \label{fig:slp-framework}
\end{figure}

As is usual in machine learning evaluation, we train and test our
classifier on two separate datasets: we must be careful that the
classifier is not trained on the same data that is used to evaluate
it. For this reason, supervised link prediction requires a train and
test framework as depicted in the bottom half of
\Cref{fig:slp-framework}. A link prediction classifier is given a set
of features related to each disconnected dyad in the period
$[t_0,t_1)$, as well as a label which indicates whether the dyad
became connected in the period $[t_1,t_2)$. From this information, it
learns a model which relates the dyad features to the probability that
a previously disconnected dyad becomes connected.  To measure the
accuracy of a link predictor, we then create a set of test dyad
features in the interval $[t'_0,t'_1)$ and use those to score the test
labels in the interval $[t'_1,t'_2)$. We measure how accurately the
scored dyads predict the set of test labels using the area under the
precision-recall curve (AUPR).

We note that the AUPR of a link predictor can fluctuate greatly: as
the behavior of users changes, so does the accuracy of the link
predictor. In order to better estimate the typical AUPR attained by a link
predictor, we can run this procedure many times; we will refer to each
run of the procedure outlined in the bottom panel of
\Cref{fig:slp-framework} as a \emph{realization}. As shown in the top
panel of \Cref{fig:slp-framework}, we shift realizations such that the
AUPR of each realization is measured using a distinct set of events.

\citet{ll-npmlp-10} convincingly argue that the link prediction problem should be
stratified over different geo\-de\-sic distances (i.e., path lengths). That is, the
disconnected dyads with a geodesic distance of $N = 2, 3, 4, \ldots$ should
each be put into different bins, and a separate classifier should be trained on
each bin. This stratification leads to better performance because the decision
boundary for each bin can be quite different, with local features (such as common
neighbors) being of primary importance at small distances, and global features
(such as preferential attachment) becoming more important at larger
distances. Thus, by treating each distance as a separate classification task, not
only does performance improve, but one can also gain insight into the particular
strengths and weaknesses of a predictor.  We therefore follow suit and
treat each distance as a separate link prediction task.


\section{Learning with Vector Clocks}
\label{sec:vector-clocks}

Having introduced the necessary background on network event data and link prediction,
we now explain how fine-grained temporal information can be exploited,
using the concept of vector clocks.

\subsection{Traditional Vector Clocks}

Vector clocks were conceptually defined in \cite{f-tmpsppo-88} and
\cite{m-vtgsds-89} as a means to track causality in concurrent systems,
but had implicitly been used before, e.g., in \citep{petal-dmids-83}, with
underlying foundations attributed to \cite{l-tcoeds-78}; for an
introduction to vector clock systems in distributed computing refer to
\citep{br-fdcptvcs-02}.
\citet{kkw-sipscn-08} brought the concept to social
network analysis by substituting message-exchanging processes with communicating
individuals. In this special setting, the basic motivation of
vector clocks is to keep track of the lower bound of
how out-of-date a person is with respect to every other person at any time,
assuming that information spreads according to a given time-ordered list of
communication events.

\begin{figure}[t]
  \centering
  \includegraphics[width=.48\textwidth]{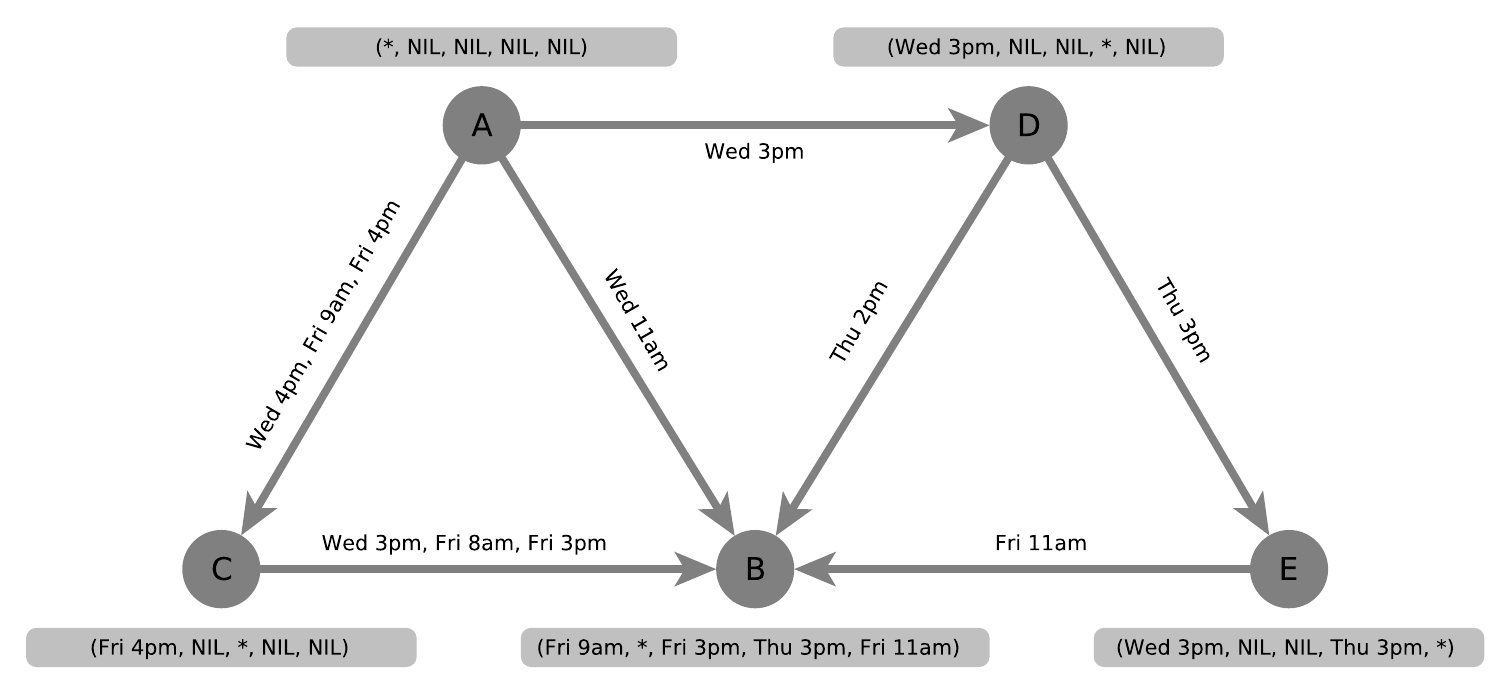}
  \caption{\label{fig:vcexample} The basic idea of vector clocks: each
    node's vector clock (in grey) keeps track of the most recent
    information it could have on the other actors in the network.}
\end{figure}

Reconsider the trip-planning example we introduced in
\Cref{sec:motivation}. There we asked: which nodes could possibly know
about the most recent details, through either direct or indirect
updates? Vector clocks provide a way of answering this question by
keeping track of the last possible update that a node could have
received from each other node: the vector clock (grey box) next to
node~E in \Cref{fig:vcexample} indicates that~E
cannot possibly have received information from~A more recent
than Wednesday at 3pm, that it could have received no information
whatsoever from nodes~B and~C, and so on. Each node
is always assumed to have up-to date information on itself.

Formally, let $(t_i,s_i,r_i)$, $i\in\mathbb{N}$, a sequence of
(time,sender,re\-ceiv\-er) tuples satisfying $t_i\leq t_{i+1}$ and
$s_i\not=r_i$.  The set of individuals is defined implicitly by
$V=\bigcup_{i\in\mathbb{N}}\{s_i,r_i\}$.  At time $t_i$, sender $s_i$
and receiver $r_i$ exchange \emph{direct} information about
themselves, and \emph{indirect} information about others that result
from communication events in the past ($t<t_i$). That is, information
can not be forwarded instantaneously but with arbitrary small delay.
Now, a \emph{vector clock} is a multivariate function $\phi_{v,t} = (
\phi_{v,t}(u) : u \in V)$, in which $v$'s \emph{temporal view}
$\phi_{v,t}(u)$ on $u$ at time $t$ is defined as the time-stamp
$t^*\leq t$ of the latest information from $u$ that could have
possibly reached $v$ (directly or indirectly) until time $t$.  At any time, each
actor is up-to-date with respect to itself,
$\phi_{v,t}(v)=t$. Temporal views on others can be tracked online as
step functions resulting from component-wise maximum calculations of
$\phi_{s_i,t_i}$ and $\phi_{r_i,t_i}$ at time-steps $t_i$.
Intuitively, $\phi_{s_i,t_i}$ is updated if and only if a
communication event is mutual (such as telephone conversations or
meetings), while the update is restricted to $\phi_{r_i,t_i}$ if a
communication event is directed (such as email-, text-, or
Twitter-messages).

%

A major drawback of traditional vector clocks is poor scalability. This results from
quadratic space requirements to maintain complete temporal views, along with
efficiency problems when performing linear-size maximum calculations on every
event.  Therefore, traditional vector clocks are too expensive to maintain and manipulate in
large graphs.  While quadratic space and linear bandwidth requirements are
necessary to allow for exact calculations in the general case
\citep{cb-cslcds-91}, approximate calculations of a limited number of temporal
views \citep{tra-pccslcds-99} and less expansive update algorithms in restricted
settings, such as acyclic communication graphs \citep{msv-elmpc-91}, have been
proposed. For suitable topologies, additional data structures can be used to reduce 
the bandwidth of information to be forwarded \citep{sk-eivc-92}.

\subsection{Social Vector Clocks}

In contrast to those enhancements stemming from the literature on
distributed computing, we propose a modification that is tailored to
\emph{social} communication networks. While the original formulation
of vector clocks is interesting for social networks because it
captures the process of gossip and indirect communication, it does so
in an exaggerated and almost clumsy manner. Experiments
\citep{k-swpap-00} on the small-world property of social networks
\cite{m-swp-67,ws-cdswn-98} and the investigation in
\citep{kkw-sipscn-08} suggest that, in the vector clock update
algorithm described above, nodes will soon receive huge amounts of
information on people they have never met, and whom even their own
contacts have never interacted with directly. Indeed, in our own
initial experimentation, we found that most actors quickly attain a
non-null temporal view with most other actors in the system, and that single
communication events often cause an actor to be updated on nearly all of
the other actors.

These global updates are hard to justify because they do not seem to resemble
social communication. In other words, while exchanging system-wide
information is important in the context of distributed computing, such
massive information exchanges do not occur when two people communicate
with each other. Rather than updating each other on most of the other
actors in the system, the nature of social communication is
\emph{bounded} by cognitive limits; such as the number of
acquaintances, which does not scale with the size of the overall
population \cite{hd-snsh-03}.

We observe that when two people meet and talk about third parties,
they are likely to discuss mutual acquaintances, or at least restrict
the conversation to people at least one of them has met directly.
Compared to this circle of acquaintances and mutual acquaintances,
they are relatively unlikely to talk about any given friend of a
friend of a friend, whom neither knows directly. Based on this
observation, we propose to bound the reach of indirect updates.  Not
only does this make the vector clock update process more closely
approximate how indirect updates actually take place in social
communication, but in practice this restriction also substantially
reduces the memory used by the algorithm, making it scale to
large \emph{sparse} social networks with millions of actors and
billions of communication events.

Our modification adds one parameter $\mu$ to the vector
clock framework, which restricts how far information can travel along
time-respecting paths; we will also refer to this parameter as the
\emph{reach} of indirect updates. More precisely, the
reach of indirect information is bounded by the minimal number
of hops it ever took a chunk of information to travel between a pair
of actors on time-respecting paths. Consider the consequence of
assigning the following values to $\mu$:
\begin{description}
\item[$\mu=1$] restricts the creation of temporal views to those pairs
  of actors that have already communicated directly: A node $r$ can
  receive an indirect update on a node $u$ if and only if this
  \emph{receiver} $r$ has previously had a direct update from $u$.
\item[$\mu=2$] additionally allows the creation of temporal views for
  dis\-tance-two neighbors (where distance is measured using
  time-re\-spect\-ing paths).  That is, when considering whether node $r$
  can receive an indirect update on a node $u$ via a direct update
  from node $s$, it is always sufficient that the \emph{sender} $s$
  previously had a direct contact with $u$. We note that this case has
  been shown to be especially important in information brokerage
  \citep{b-sbeilsmba-07}.
\item[$\mu=\infty$] corresponds to the classical vector clock
  algorithm with unlimited information spread, and in practice quickly
  results in quadratic space requirements.
\end{description}
This modification is
straightforward to implement, because using the vector-clock
framework, it is trivial to track the length of shortest
time-respecting paths: When processing a communication event
$(t,s,r)$, the minimum number of hops it ever took a chunk of
information about $u$ to reach $r$ is given by
\begin{equation*}
\textnormal{dist}_{t_i}(u,r)=\min(\textnormal{dist}_{t_{i-1}}(u,s)+1,\textnormal{dist}_{t_{i-1}}(u,r)),  
\end{equation*}
where $\textnormal{dist}_{t}(a,b)$ refers to the length of the
shortest time-respecting $(a,b)$-path until time $t$.  In this way,
distances are directed, respecting the ordering of events, and
decreasing over the course of time. In our implementation, distances
are not known to the source of an information chain, but saved
together with the vector clock of the target.  Once a short
information chain has been observed, a corresponding temporal view is
established and also allowed to be updated by longer information
chains.  In practice, this modification reduces memory requirements
very substantially.




\begin{table*}[t]
  \centering
  \begin{tabular}{lrrrrrrr}
    \toprule[1.1pt]
      & & \multicolumn{2}{c}{$N=2$} & \multicolumn{2}{c}{$N=3$} & \multicolumn{2}{c}{$N=4$} \\
      \cmidrule(lr){3-4} \cmidrule(lr){5-6} \cmidrule(lr){7-8}
      Dataset & Realizations & Avg. Pos. & Avg. Neg. & Avg. Pos. & Avg. Neg. & Avg. Pos. & Avg. Neg.\\
      \midrule
      \election & 26 & 796 & 243658 & 186 & 922854 & 41 & 1373586\\
      \electionExt & 26 & 1664 & 516415 & 583 & 1973244 & 198 & 2888931\\
      \olympics & 12 & 182 & 15704 & 38 & 40257 & 10 & 42641\\
      \irvine & 4 & 478 & 167674 & 675 & 536188 & 95 & 348814\\
      \studivz & 23 & 1332 & 751172 & 706 & 5765692 & - & - \\
      \bottomrule[1.1pt]
    \end{tabular}
    \caption{Link prediction statistics by dataset and path length
      $N$. These values report the average over all realizations of each
      experiment. For each realization, measurements are based on the graph associated with the
      interval $[t'_0,t'_1)$ and the new edges formed in the interval $[t'_1,t'_2)$.}
    \label{tab:lp-stats}
\end{table*}


\subsection{A Link Predictor with Social Vector Clocks}
\label{sec:vc-features}

We have described how social vector clocks can, in real time, keep
track of the most recent information that could have possibly traveled
between pairs of nodes. While we compute the social vector clocks, we can
easily derive several features that may be useful for link
prediction. These features can then be combined using a supervised
link predictor as we outlined in \Cref{sec:supervised-lp}.

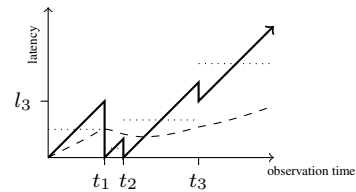
\begin{figure}[b]
   \centering
    \begin{tikzpicture}
      \draw[color=white] (-1,0) -- (4,0); \draw[->] (0,0) -- (3,0); \draw[->]
      (0,0) -- (0,2);
      \draw (3.5,0) node [below] {\tiny observation time};
      \draw (0,1.5) node [above,rotate=90] {\tiny latency};

     \draw[thick,->] (0,0) -- (0.75,0.75) -- (0.75,0) -- (1,0.25) -- (1,0) -- (2,1) -- (2,0.75) -- (3,1.75); 

     \draw[] (0.75,0) -- (0.75,-0.1); \draw[] (1,0) -- (1,-0.1); \draw[] (2,0) --
      (2,-0.1); \draw[] (0.75,-0.1) node[below] {\small $t_1\ $}; \draw[]
      (1,-0.1) node[below] {\small $\ \ t_2$}; \draw[] (2,-0.1) node[below]
      {\small $t_3$};

      \draw[] (0,0.75) -- (-0.1,0.75); \draw[] (-0.1,0.75) node[left] {\small
        $l_3$};
        
        
        \draw[dotted] (0,0.75/2) -- (0.75,0.75/2);
        \draw[dotted] (0.75,0.25/2) -- (1,0.25/2);
        \draw[dotted] (1,0.5) -- (2,0.5);
        \draw[dotted] (2,1.25) -- (3,1.25);


        \draw[dashed] plot [smooth] coordinates {(0.0,0.0)(0.25,0.125)(0.5,0.25)(0.75,0.375)(1.0,0.313)(1.25,0.275)(1.5,0.292)(1.75,0.339)(2.0,0.406)(2.25,0.458)(2.5,0.525)(2.75,0.602)(3.0,0.688)};
        
    \end{tikzpicture}
  \caption{\label{fig:tracking}Deriving link prediction features from social vector clocks: 
  Example of a dyad with \emph{two direct updates} (at time $t_1$ and $t_2$; the new latency becomes zero),
  followed by \emph{one indirect update} (at time $t_3$; the new latency is $l_3>0$).
  The \emph{current latency} (solid line) is a linear jump function resulting from $t-\phi_{v,t}(u)$. 
  The \emph{expected latency} (dashed line) is the weighted mean of average latencies (dotted horizontal lines) between vector clock updates.
  }
\end{figure}

A first feature is immediately derived from the temporal views that are saved in the vector clocks:
the \emph{current latency} is defined as the difference between the current time and the timestamp saved in the temporal view.
As second and third features, we track the number of
\emph{direct updates} and \emph{indirect updates} that occur between a
pair of actors as the vector clocks are computed.  
A fourth feature we calculate is the \emph{expected latency} between each pair of nodes, 
which can be thought of as the best guess on how out-of-date an actor is about another at any point
in the observation window. See \Cref{fig:tracking} for an illustration
of all these features.


Some users of a service like Twitter may be much more active than
others. This heterogeneity will mean that some users will have a
latency of weeks with their closest contacts, whereas others will
typically have a latency of days or hours with their closest
contacts. Such heterogeneity may make it hard for a 
classifier to detect a decision boundary; for this reason, in addition
to keeping track of the absolute values of the current and expected
latencies, we keep track of the ranks. That is, from the
perspective of a given node $i$, we sort each of $i$'s temporal views by
their current latency, and then rank the corresponding dyads---this yields an
additional feature for each dyad $\{i,j\}$. We do the same for the
expected latency.

So far we have described six features for each dyad: the current
latency (both absolute value and rank), the expected latency (absolute
value and rank), the number of direct updates, and the number of
indirect updates. All of these features can be kept track of in real
time and in practice they add little computational overhead. In the
context of directed link prediction, for each directed dyad we can
keep track of all six of these features in both directions, yielding
twelve features. Finally, our definition of social vector clocks
included one parameter~$\mu$ which bounds how far information can
travel.  In practice one might not know which value of this parameter
will lead to the best results; in such a case, one can simply run
multiple instances of the vector clocks in parallel, each with a
different value of the reach parameter, and combine all the resulting
features. A classifier can then learn which feature set is the
most informative. For example, in our evaluation below, we run \alg
with three different reach parameters: 1, 2, and $\infty$, which
creates a total of 36~features for each directed dyad.
 

\section{Evaluation \& Results}

\subsection{Datasets}
\label{datasets}
Two of the datasets we consider come from Twitter. While Twitter is
often used as a medium for impersonally broadcasting messages to large
numbers of followers, it also supports more targeted forms of
communication, in which users explicitly refer to each other.  This
targeted (although public) communication occurs in the form of
retweets (in which one user rebroadcasts another users tweet, and
attributes the tweet to its source) and user mentions, where the $@$
symbol is used to explicitly refer to a user.  In the Twitter data
that we analyze here, we filter Twitter datasets to include only this
targeted form of communication (i.e., those with retweets or user
mentions). We remove self loops, and if a tweet mentions more than one
user, we turn it into as many events as there are users mentioned in
the tweet.  With this representation, the data corresponds to the
basic scenario of dyadic communication event streams underlying our
investigation: we are given a sequence $(t_i,s_i,r_i)$,
$i\in\mathbb{N}$, of (time,sender,re\-ceiv\-er) tuples satisfying
$t_i\leq t_{i+1}$ and $s_i\not=r_i$.

\begin{description} 
 \item[Twitter UK Olympics Data]
 The \olympics dataset covers Twitter communication among
a set of 499 UK Olympic athletes over the course of the 18 months leading up
to the 2012 Summer Olympic Games, including 730,880 tweets.
It was introduced in \cite{g-ittcula-12} and is based on
a list of UK athletes curated by \textit{The Telegraph}.\footnote{{\footnotesize\url{twitter.com/#!/Telegraph2012/london2012}}} We remove all
tweets that are not user mentions or retweets between this core set of 499 users,
a step which reduces the dataset to 93,613 tweets among 486 users.

\item[Twitter US Elections Data]
Similar to the \olympics dataset, the \election dataset is based on a curated list
of Twitter users, in this case curated by Storyful, a commercial news gathering
platform targeted at journalists. One of Storyful's features is topical Twitter
lists, which journalists can subscribe to in order to remain well informed on a
given topic. Here we scraped tweets by users on Storyful's US 2012 Presidential
election list. In addition we added the Twitter accounts of all those candidates
seeking office at the level of governor, US Senator, or member of the
US House of Representatives. In total, this dataset covers the date range from
Jan.\,1st 2012 to Nov.\,9th, 2012, and includes 392,662 user
mentions and retweets among 2447 Twitter accounts.
Additionally, we created an extended version of this data\-set, which we will refer
to as \electionExt, by collecting the
tweets associated with Twitter users who were often mentioned in \election. This
extended dataset includes 546,329 tweets among 5,632 users over the same date
range as \election.

\item[StudiVZ Wall posts]
StudiVZ was created in~2005 as a German competitor for international online social networks. For several
years it was the most popular online social network in Germany, although it has
recently been overtaken by Facebook. The \studivz dataset we examine
here is based on a crawl of a single university's subnetwork; it is described
in~\cite{lbs-iosnsim-11}. While in~\cite{lbs-iosnsim-11} the static friendship
network is analyzed, here we focus on the wall post (``Pinnwand'') data. The dataset is the
largest we look at here, containing 26,180 nodes and 886,241 events.

\item[UC Irvine]
\Citet{poc-naoc-09} introduced an event-based dataset which comes from a social
networking site set up for the students of the University of California at
Irvine. Each event in this dataset is a message---it is unclear whether these are
private or public messages. While the dataset covers a period from April to
October~2004, the great majority of the events occur between mid April and mid
June. Starting in mid June, there is a two week period in which no events occur,
and for the remainder of the dataset very few messages are sent. For this reason,
we look only at the period from April~10th to June~15th. 
\end{description}

\subsection{Experiment Setup}
\label{eval-setup}

The ``high-performance'' link predictor (HPLP+) introduced in~\cite{ll-npmlp-10}
is a state of the art link predictor which combines some of the strongest unsupervised link predictors.
In the following experiments, HPLP+ acts as the baseline predictor
and our objective is to evaluate the performance of the vector clock link predictor (\vclp) described in \Cref{sec:vc-features}, 
and a combined predictor which uses the features from both \vclp and HPLP+.\footnote{We use the LPMade link prediction framework to
compute HPLP+; this is the author's reference implementation~\cite{l-lplpme-11}.}

Framing a supervised link prediction task requires several parameters. One
important parameter is the choice of classifier: as in~\cite{ll-npmlp-10}, we
used bagged forests, a technique suited for the extremely imbalanced classes
found in link prediction.  However, rather than bagging ten random forests, we
bag ten Stochastic Gradient Boosting classifiers. We use the implementation
provided in the scikit-learn python package~\cite{scikit-learn}, using 1000~trees
in each classifier, setting the learning rate to~0.005, and subsampling rate to~0.5.
In each bag we sampled from the positive instances
with replacement, and undersampled from the negative instances with replacement
such that the class imbalance ratio was 10 negative for every positive.

Another important experimental parameter is whether the link prediction is
directed or undirected. In all of our data\-sets,
edge direction is highly relevant---for example, I might mention President Obama
in a tweet, but Obama mentioning me in a tweet would have a completely different
meaning. For this reason, we restrict our evaluation to
directed link prediction.
As one can see in \Cref{tab:lp-stats}, as the directed geodesic distance ($N$) 
in the link prediction task increases, 
classes become severely imbalanced, and in
the case of \olympics, hardly any new links form. In \olympics in general, the
classifier has very little positively labeled data to train on, which increases
the risk of overfitting when extra features are added.

We must also specify some parameters related to the width of the
temporal windows used in the evaluation. In principle, we wanted to
make the duration of training period long enough so that a clear and
stable snapshot of the network has emerged, and then evaluate on
events that occur just after the end of the training
period. Therefore, wherever possible we used a training window of 120
days and a test window of 7 days. (In other words, the width of the
red bars in \Cref{fig:slp-framework} is 120 days and the width of the
blue bars is 7 days.) However, given the short duration of the UC
Irvine dataset, we use a shorter training period than in the other
datasets, and are not able to run as many realizations of our
evaluation; we set the training period to 28 days. Furthermore, the
small size of \olympics meant that in the seven day test period very
few new links emerged, leaving the classifier with too little data to
train on. Thus, for \olympics we set the test width to 14 days.

\subsection{Experiment Evaluation}
\label{eval-principle}
\begin{figure}[ht!]
  \centering
  \includegraphics[width=0.80\linewidth]{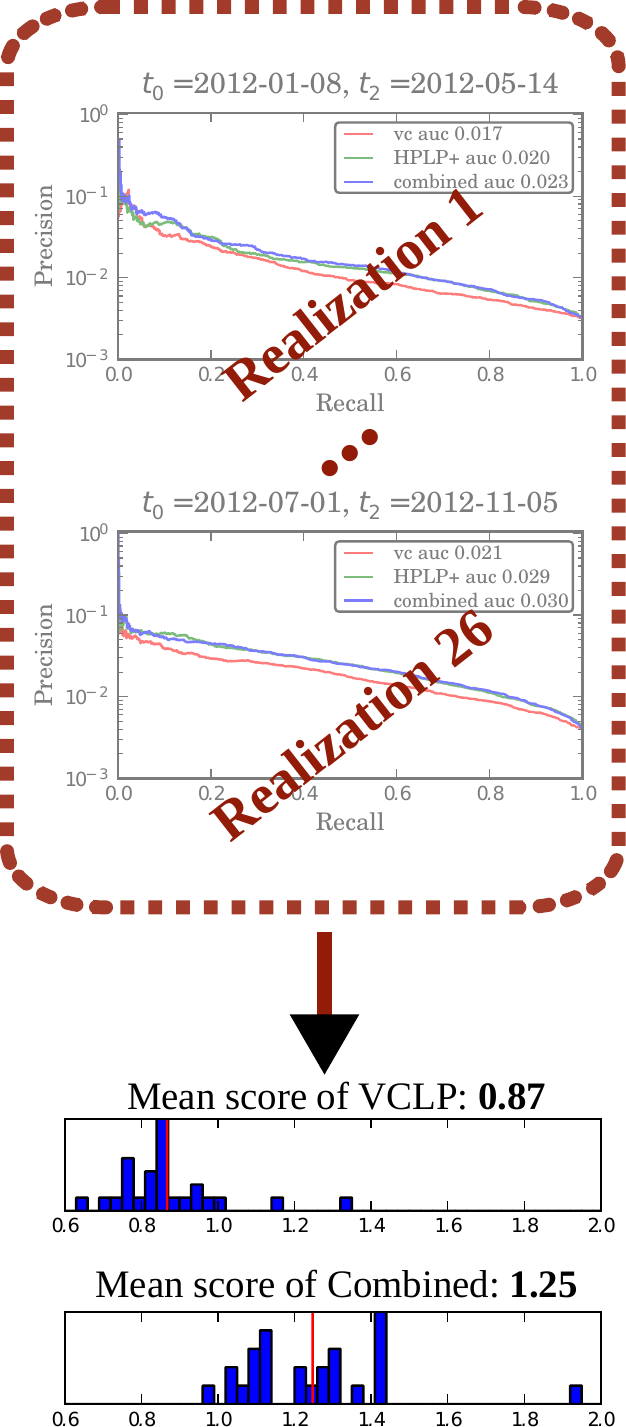}
  \caption{
  An overview of how we scored the link prediction task. Link
    predictors are first run on each realization of the experiment. 
    In each realization, precision-recall curves are constructed. The
    AUPR for each predictor is then measured, and the AUPR of \vclp and the
    combined link predictor is then divided by the AUPR of HPLP+; this score
    is the relative performance of each predictor with HPLP+ as the baseline. 
    For each dataset, the average of these scores is then reported;
    stratified over different geo\-de\-sic distances.
    }
  \label{fig:eval-overview}
\end{figure}

The number of realizations performed on each dataset is indicated in
\Cref{tab:lp-stats}.  For each realization, we record the \prc of each
predictor, leaving us with a sequence of pre\-ci\-sion-recall plots such
as those presented in the upper section of
\Cref{fig:eval-overview}. We are interested in how \vclp and the
combined predictor perform relative to HPLP+, so we summarize them as
follows (as outlined in \Cref{fig:eval-overview}): We treat the
performance of HPLP+ as the baseline, and in each plot, we measure the
area under HPLP+'s \prc. We then record the area under the
precision-recall curve (AUPR) of both VCLP and the combined predictor
as a fraction of HPLP+'s AUPR. Thus, if in one realization HPLP+'s
AUPR is $0.020$ and the combined predictor's AUPR is $0.024$, then we
record the combined predictor's score as $1.2$. After recording this
score for all realizations, we are left with a distributions of scores
as in the histogram in \Cref{fig:eval-overview}. By taking the average
of these scores, we can characterize in a single number how much
better or worse \vclp and the combined predictor perform than
HPLP+. We report these averages for each experiment in
\Cref{tab:results-1}; see next section for discussion.

\begin{table*}[t]
  \centering
  \caption{Average performance of supervised link predictors relative to HPLP+.}
  \label{tab:results}
  \begin{subtable}[t]{1.0\linewidth}
    \centering
    \begin{tabular}{rlccccccccccccccc}
      & & \multicolumn{3}{c}{\election} & \multicolumn{3}{c}{\electionExt} & \multicolumn{3}{c}{\olympics} & \multicolumn{3}{c}{\irvine} & \multicolumn{3}{c}{\studivz}\\
      \cmidrule(lr){3-5} \cmidrule(lr){6-8} \cmidrule(lr){9-11} \cmidrule(lr){12-14} \cmidrule(lr){15-17}
       & $N$ & $2$ & $3$ & $4$ & $2$ & $3$ & $4$ & $2$ & $3$ & $4$ & $2$ & $3$ & $4$ & $2$ & $3$ & $4$  \\
       \cmidrule(lr){3-5} \cmidrule(lr){6-8} \cmidrule(lr){9-11} \cmidrule(lr){12-14} \cmidrule(lr){15-17}
      \vclp  &  & 0.87 & 0.86 & 1.05 & 0.92 & 0.75 & 0.86 & 0.97 & 1.07 & 1.06 & 1.11 & 1.87 & 1.12 & 1.17 & 1.20 & -\\
      Combined & & 1.25 & 1.15 & 1.00 & 1.43 & 1.38 & 1.15 & 1.08 & 1.10 & 1.06 &
      1.33 & 1.65 & 1.13 & 1.39 & 1.22 & - \\
    \end{tabular}
    \vspace{0.2cm}
    \caption{Results on predicting all edges at different distances ($N=2,3,4$); see \ref{sec:eval-results} for discussion.}
    \label{tab:results-1}
  \end{subtable}

  \vspace{0.5cm}

  \begin{subtable}[t]{1.0\linewidth}
    \centering
    \begin{tabular}{rlccccccccccccccc}
      & & \multicolumn{3}{c}{\election} & \multicolumn{3}{c}{\electionExt} & \multicolumn{3}{c}{\olympics} & \multicolumn{3}{c}{\irvine} & \multicolumn{3}{c}{\studivz}\\
      \cmidrule(lr){3-5} \cmidrule(lr){6-8} \cmidrule(lr){9-11} \cmidrule(lr){12-14} \cmidrule(lr){15-17}
       & $N$ & $2$ & $3$ & $4$ & $2$ & $3$ & $4$ & $2$ & $3$ & $4$ & $2$ & $3$ & $4$ & $2$ & $3$ & $4$  \\
       \cmidrule(lr){3-5} \cmidrule(lr){6-8} \cmidrule(lr){9-11} \cmidrule(lr){12-14} \cmidrule(lr){15-17}
      \vclp    & & 0.75 & 0.78 & 0.99 & 0.82 & 0.65 & 0.79 & 0.75 & 1.02 & 1.23 & 0.95 & 0.79 & 1.00 & 0.49 & 0.57 & - \\
      Combined & & 1.17 & 1.05 & 1.00 & 1.38 & 1.34 & 1.17 & 1.06 & 1.01 & 1.00 & 1.19 & 1.23 & 1.03 & 1.11 & 1.04 & -\\
    \end{tabular}
    \vspace{0.2cm}
    \caption{Results on predicting non-reciprocal edges at different distances; see \cref{sec:eval-reciprocity} for discussion.} 
    \label{tab:results-2}
  \end{subtable}
\end{table*}

\subsection{Results}
\label{sec:eval-results}

In \Cref{tab:results-1}, we see that VCLP on its own
performs comparably to HPLP+. Considering that HPLP+ combines 
a broad range of sophisticated graph features, 
we were surprised to see VCLP perform similarly.
Moreover, all network statistics employed by the proposed VCLP can be
kept track of online, directly on the list of communication events,
while many of the statistics included in HPLP+ have to be recalculated
whenever new links are added to the network.
Consequently, our results suggest that
link prediction with vector-clock statistics can be performed much more efficiently in 
any situation, where the model parameters are learned beforehand
and applied in real-time on a growing sequence of ``test'' events.

When the features of VCLP and HPLP+ are combined, the performance increase over
HPLP+ is substantial. In general, the performance gain is largest when
we are predicting links on dyads that have a geodesic distance
$N=2$. Performance gain decreases for greater $N$, suggesting that VCLP features are most
useful for predicting local links rather than long-range links (\irvine is an
exception to this trend, where $N=3$ sees by far the largest boost to performance).
The improvement is smallest on \olympics, perhaps because the classifier
struggles with the small number of positive training examples.

Given that stand-alone VCLP and HPLP+ yield similar prediction accuracy, 
it is interesting to observe the added value of combing both predictors.
In other words, there appear to be qualitative differences in 
the network effects that can be captured in the VCLP and HPLP+ framework.

\subsection{Controlling for reciprocity}
\label{sec:eval-reciprocity}
Imagine we're trying to predict whether a node~A will soon send its
first message to~D.  One of the features included in \vclp is ~D's
current latency with~A through direct updates~-- in other words, how
many seconds have elapsed since ~D sent a message to~A. Given the
significance of reciprocity, this feature will be extremely useful for
cases where~D has just sent a message to~A. It could be the case
that this feature alone~-- which is trivial to keep track of without
vector clocks~-- is responsible for all of the benefit that comes from
\vclp. In that case, we could simply keep track of this single feature
and forget about vector clocks.

\begin{figure}[h]
  \centering
  \includegraphics[width=0.9\linewidth]{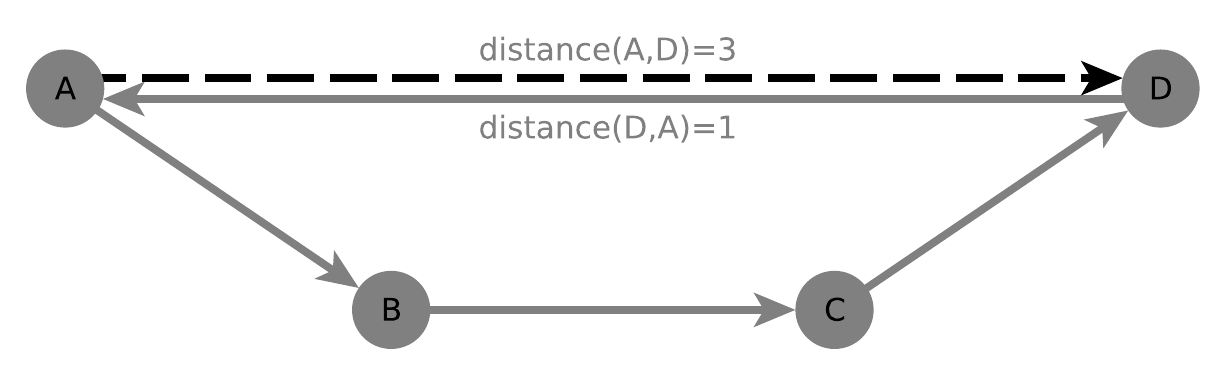}
  \caption{The directed dyad (A,~D) has a geodesic distance of 3, but the
    distance of dyad (D,~A) is 1. This dyad would be included in the experiments
    whose results are presented in \Cref{tab:results-1}, but would be excluded from
    the experiments whose results are presented in \Cref{tab:results-2}.}
  \label{fig:directed-distance}
\end{figure}

To measure whether this is the case, we run the entire evaluation
again, but exclude all dyads where~D has had any direct contact
with~A; see \Cref{fig:directed-distance}. The results presented in
\Cref{tab:results-2} are in the same units as the results presented in
\Cref{tab:results-1}. We observe that the performance of \vclp does
indeed drop, but that there is still a significant benefit provided by
combining the features of \vclp and HPLP+. Again, we stress that the
lackluster performance on \olympics may be due to the small number of
new links that form, which provides very few positive training
examples.

\section{Summary}

The current approach used by state of the art link predictors is to operate in a
panel data setting, in which finer-grained temporal information is ignored.
In cases where link formation is not driven by cascades of information,
such an approach might be appropriate. Regarding co-authorship networks, for instance,
precise information on the sequence and spacing of
events may be largely irrelevant or even misleading, and so it may be reasonable to
aggregate away information on exact publication dates.
However, in some networks~-- such as the Twitter retweet/mention networks
mentioned here~-- information cascades are an important mechanism for driving the
formation of edges. In such a setting, the information contained in the exact
sequence and spacing of events is highly relevant, and so the approach commonly
employed in link prediction~-- to simply aggregate event data into panel data~-- is highly
questionable.  For example, the mechanism of reciprocity has
been shown to be important in the context of directed link prediction. Thus, if
we are trying to predict whether~A will send a message to~B, then an
extremely useful piece of information is whether~B has recently sent~A a
message; if so, it is likely that~A will respond to~B. By aggregating all
events into a static graph, traditional link prediction schemes cannot exploit
such simple and useful mechanisms.

Our results suggest that dyadic features that exploit fine-grained temporal
information beyond reciprocity are highly relevant for predicting which actors
will communicate for the first time in the near future.  The approach we
introduce here, called the Vector Clock Link Predictor (VCLP), is based on
keeping track of the latencies between all presumably relevant pairs of actors.
The basic idea is to exploit information on how out of date a
node~A is with respect to another node~B, and for doing so we adopt the
concept of vector clocks.  As an essential modification, we parameterized the traditional
vector-clock concept to bound the reach of indirect information.  Not only does
this make the vector-clock update process more closely approximate how indirect
updates actually take place in social communication, but in practice this
restriction also dramatically reduces the memory used by the algorithm, thus
making it applicable to large \emph{sparse} social networks with millions of
actors and billions of communication events.

We have demonstrated that binary classifiers can indeed exploit actor latencies
to improve accuracy in link prediction. Even HPLP+, a classifier which utilizes a
wide range of graph features based on aggregated panel data, can perform
substantially better when provided with additional features based on vector
clocks.  Moreover, VCLP on its own already performs comparably to HPLP+, which allows for
much more efficient link prediction in any situation where the model parameters
are learned beforehand and applied in real-time on a growing sequence of events.

Both of the supervised link prediction schemes considered here are
based on many features, and by adding or removing various features
many variations of social vector clocks are conceivable.  Even with
the intuitive motivation for social vector clocks and their
demonstrated performance, we have not necessarily advanced the
understanding of the actual mechanisms behind link formation.  We are
keen to gain more detailed insight into the link prediction problem
for specific types of interaction, e.g., by combining
feature-selection schemes and more elaborate substantive theories.



\end{document}